\documentclass[journal]{IEEEtran}
\usepackage{cite}
\usepackage{color}
\usepackage{multirow}
\usepackage{booktabs,multirow,array}
\usepackage[table]{xcolor} 
\usepackage{ltxcmds}
\usepackage{footnote}
\usepackage{amsthm}

\usepackage[table]{xcolor}
\usepackage{subfigure}
\usepackage{lipsum}

\usepackage{amssymb} 
\usepackage{pdflscape}
\usepackage{afterpage}

\usepackage{pifont}
\ifCLASSINFOpdf
  \usepackage[pdftex]{graphicx}
\else
  \usepackage[dvips]{graphicx}
\fi
\begin{document}
\title{A Fog-Based Security Framework for Large-Scale Industrial Internet of Things Environments}

%

\author{Hejia Zhou,
        Shantanu Pal,
        Zahra Jadidi,
        and~Alireza Jolfaei
        
\thanks{H. Zhou is with the School of Computer Science, Queensland University of Technology, QLD 4000, Australia (e-mail: hejia.zhou@connect.qut.edu.au)}
\thanks{S. Pal is with the School of Information Technology, Deakin University, Melbourne, VIC 3125, Australia (e-mail: shantanu.pal@deakin.edu.au)}
\thanks{Z. Jadidi is with the School of Information and Communication Technology, Griffith University, Gold Coast Campus, QLD 4222, Australia (e-mail: z.jadidi@griffith.edu.au)}
\thanks{A. Jolfaei is with the College of Science and Engineering, Flinders University, Adelaide, SA 5042, Australia (e-mail: alireza.jolfaei@flinders.edu.au)}
}

%
%

\markboth{}%
{Author\MakeLowercase{\textit{Zhou et~al.}}: Internet of Things}
%



\maketitle


\begin{abstract}
The Industrial Internet of Things (IIoT) is a developing research area with potential global Internet connectivity, turning everyday objects into intelligent devices with more autonomous activities. IIoT services and applications are not only being used in smart homes and smart cities, but they have also become an essential element of the Industry 4.0 concept. The emergence of the IIoT helps traditional industries simplify production processes, reduce production costs, and improve industrial efficiency. However, the involvement of many heterogeneous devices, the use of third-party software, and the resource-constrained nature of the IoT devices bring new security risks to the production chain and expose vulnerabilities to the systems. The Distributed Denial of Service (DDoS) attacks are significant, among others. This article analyzes the threats and attacks in the IIoT and discusses how DDoS attacks impact the production process and communication dysfunctions with IIoT services and applications. This article also proposes a reference security framework that enhances the advantages of fog computing to demonstrate countermeasures against DDoS attacks and possible strategies to mitigate such attacks at scale. 
\end{abstract}


%
\IEEEpeerreviewmaketitle

\section{Introduction}
\label{introduction}
The Industrial Internet of Things (IIoT) triggered the fourth industrial revolution (known as Industry 4.0) based on Cyber Physical Systems (CPS), which defines various use cases from connected digital technology, mobile cloud computing and Internet of Things (IoT) to promote the efficiency, effectiveness, support for varied data, higher manufacture, mechanization, and wide-ranging knowledge~\cite{huo2022comprehensive}. IIoT has the potential to enhance traditional IoT communications in a new way (with a large-scale deployment), which allows all devices to share and use the same data in the same environment. Some characteristics of the IoT are~\cite{pal2021analysis}:

\begin{itemize}
    \item The IoT cannot simply be seen as an extension of the Internet. IoT is built on a unique infrastructure and will be a new set of independent systems, although some of the underlying facilities will still be dependent on existing Internet technologies.
    
    \item The IoT could be accompanied by new business development to a single network to automate operational control and optimise the workflow.
    
    \item The IoT includes a variety of different communication modes, object-to-person communication and object-to-object communication, with particular emphasis on machine-to-machine communication (M2M).
    
\end{itemize}

\begin{figure}[t]
 \centering
    \includegraphics[scale=1.72]{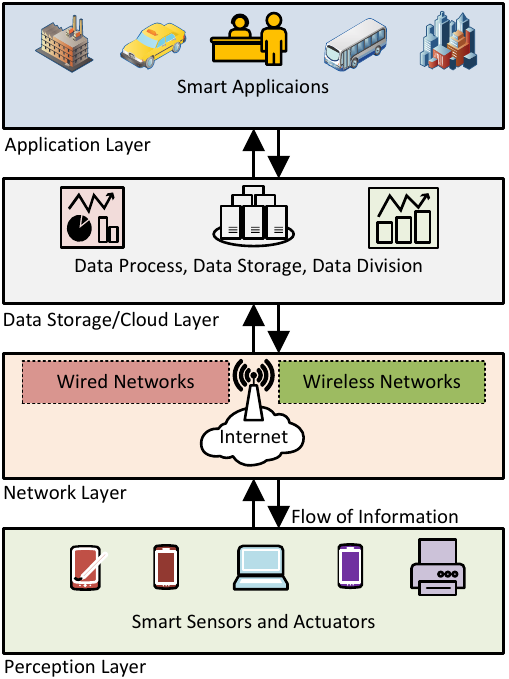}
    \caption{A simple IoT architecture consists of four layers.}
    \label{fig-1-view}
    \small
\end{figure}

Based on the characteristics mentioned above, the IoT can be seen as a distributed system connecting many information sensing devices and applications through the Internet and uniquely identifying entities within a broad network area.

The IIoT systems integrate the features of an IoT system. Therefore, it is essential to examine the basic building blocks of an IoT system. Several potential architectures discuss the layer view of an IoT system. These are composed of three to seven layers. However, in this article, we consider a basic IoT architecture consisting of four layers (as shown in Fig.~\ref{fig-1-view}). We argue that a four-layer architecture can efficiently distribute the various components of an IoT architecture. The bottom layer is called the \textit{perception layer}, which is used for sensing the physical world. The perception layer includes smart sensing devices represented by sensors, actuators, location tracking devices, e.g., GPS (Global Positioning System) and smart terminals, etc. The perception layer is the basis for the IoT to obtain information and data, and its purpose is to achieve data sensing on a large scale. The next layer is the \textit{network layer} used for data transmission. The network layer includes the access network and the core network. The network layer is the transmission layer for collected data. The collected data is stored on the \textit{data storage layer} (also known as the cloud layer) located above the network layer. Finally, the top layer is the \textit{application layer}, where the users use various data to achieve their business or manufacturing goals~\cite{lin2017survey}. 

\begin{figure*}[t]
 \centering
    \includegraphics[scale=1.75]{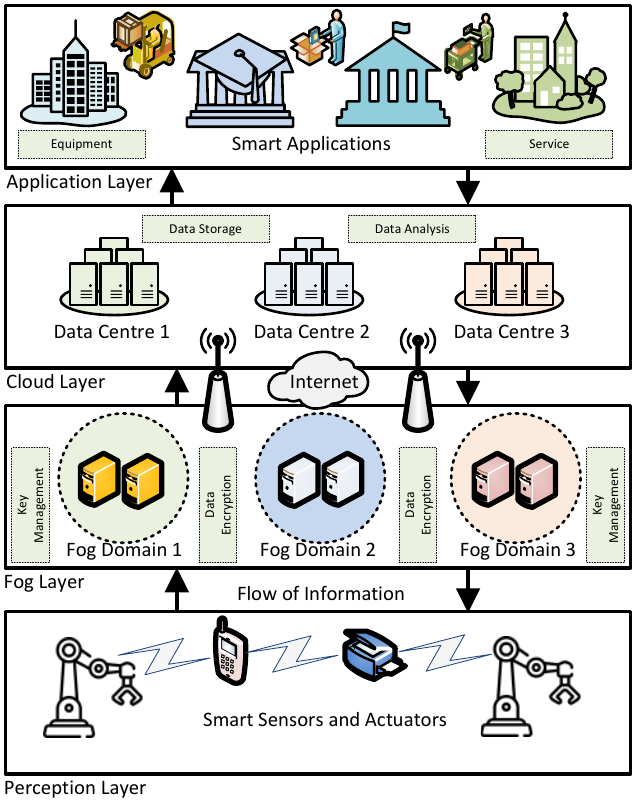}
    \caption{A simple overview of a fog-based IIoT architecture.}
    \label{fig-2-view}
    \small
\end{figure*}

A similar layered architecture can be considered for an IIoT system for efficient data processing locally and improved response time. For instance, in an IIoT system, the perception layer collects data and transmits it to the data storage layer through the network layer for further processing. The data storage layer then analyses and processes the data and sends it to the application layer to furnish users with a wealth of specific services, for example, smart transportation, smart home, smart health, and smart city. 

\textit{Roadmap of the Paper:} Next, we discuss the concept of an IIoT architecture with more granularity with the inclusion of `fog computing'. Towards this, we discuss a fog-based reference framework for IIoT systems that can efficiently address DDoS (Distributed Denial of Service) attacks at scale. We also provide a list of mitigation strategies for DDoS attacks in an IIoT system based on the discussed reference framework. We also discuss a security analysis of the proposed architecture.

\section{The Concept of Fog-Based IIoT Architecture}
\label{fog-iiot-concept}

The design of an IIoT framework incorporates the concept of a fog layer. The fog layer situates between the perception and cloud layers (i.e., the data storage layer) and is responsible for data transmission, encryption, and area management~\cite{hu2017survey}. As an extension of the cloud layer, the fog layer helps the cloud layer to take on some of the data processing, distributed area management, distributed signature authentication, data re-encryption and critical management responsibilities. The fog layer interfaces upwards to the cloud layer, where it is managed and receives data sets from the cloud layer, and downwards to the perception layer, where it regionalizes and addresses the various sensors that access the IIoT, receives data requests from devices, authorizes the requests and establishes the data transmission links. Significantly, the use of the fog layer can perform some of the basic functions of the network layer, e.g., data transmission from the perception layer to the cloud layer. 

As Fig.~\ref{fig-2-view} shows, a general IIoT architecture can consist of the basic elements of an IoT architecture but with a fog layer, which we discussed above. The IoT and the IIoT are very similar in that they are related to a network of intelligent devices, machines, and sensors. The most common difference between the IoT and IIoT is that IoT is derived from conceptual terms, e.g., creation, concatenation of acts, manipulation of posters and other instances of constructive expression. Since the IIoT is extremely forgiving of others and is very powerful, it has a high degree of language, which means that the data and ideas generated in the reproductive community are likely to be more sophisticated and sensitive~\cite{chalapathi2021industrial}.



Although we find that the layering of the IIoT architecture can be three layers, four layers or even seven layers, it is more challenging to implement and requires consideration of its security, privacy, and trust issues between these layers. Moreover, because of the instability of network protocols and less human intervention, it becomes increasingly vulnerable to various security threats in an IIoT system. The additional specialist reason is that technicians have found that the cloud-only network layer is becoming challenging to support the large number of devices accessing and transmitting data. Hence the introduction of the concept of a fog layer is pivotal. Next, we analyse the advantages of including fog layers in an IIoT architecture and demonstrate whether using them can lead to more secure and efficient information processing, enhancing storage capabilities, security, privacy, and trust.

\section{Security Challenges}
IoT and IIoT have some common characteristics, and according to those common factors, it is hard to distinguish different attacks individually for each system, especially the DDoS attacks~\cite{salim2020distributed}. Therefore, an attack that impacts an IoT system can easily affect an IIoT system. Some concerns regarding emerging security issues in IIoT systems are~\cite{pal2017design}: 


\subsubsection{Threats Built into the IIoT}
IIoT threats, including the threat of attacks on databases, are intertwined with other trends in 2021. In a world of increasing automation, many attacks are focused on supply chains and manufacturing. There are many applications of IIoT in these areas, and updating devices is not always a priority. As we experience more new types of attacks against the IIoT (for example, Mirai botnets and WannaCry), one fundamental question is whether we can update aging firmware to give it the defence it needs.

\subsubsection{Threat of Artificial Intelligence (AI) in the IIoT} In recent years, it is likely to be the challenges of AI-driven threats to the IIoT. AI-based attacks have been occurring since 2007, mainly against social engineering attacks (simulating human chatter) and augmented DDoS attacks. AI systems are better than humans at performing many elements of IoT threats, e.g., repetitive tasks, interactive responses, and processing extensive data sets. In general, AI helps attackers amplify the IoT threat. In 2021, we not only envision new AI-based IoT threats but also realize the magnitude of such attacks. We should look for the usual cyber breaches and other attacks, but they are deployed faster, on a larger scale, and more flexible, automated, and customized than in the past.

\subsubsection{Data Protection Challenges}
In an IIoT-connected world, data protection becomes significant because it can be transferred between multiple devices in seconds. It can store on a mobile device, and in the next second, it may be stored on the web and then in the cloud. All this data is transferred over the Internet, leading to data leakage due to this high mobility of data. Once the data is compromised, attackers can make unauthorized use of this sensitive data to other organizations that violate the privacy and security rights of the data (and, in the long turn, the reputation of the organizations). In addition, even if the attackers do not compromise data, service providers may not comply with laws and regulations, leading to data privacy breaches.

\section{Impact of DDoS Attacks on IIoT}


In terms of the motivations used to deploy DDoS attacks, they can be seen into the following areas~\cite{bhardwaj2021distributed}:

\begin{itemize}
    \item \textit{Abuse of a Reasonable Request for Service:} The attacker excessively requests the normal services of the system, taking up too many service resources and causing the system to be overloaded. These service resources include network bandwidth, file system space capacity, and a number of open processes or connections.
    
    \item \textit{Creating High Volumes of Unwanted Data:} Maliciously creating and sending large amounts of random and useless data packets to occupy network bandwidth with this high volume of useless data, causing network congestion.
    
    \item \textit{Exploiting Flaws in the Transport Interest Protocol:} Constructing and sending malformed packets that cause errors or crashes that cannot be processed by the target primary server.
    
    \item \textit{Exploit a Vulnerability in the Service Program:} This behavior targets a specific vulnerability in the service program on the host and sends some targeted data in a special format that causes the service to process it incorrectly and deny service.
\end{itemize}

Compared with other security attacks, DDoS attacks are more concealed and harmful due to the resource-constrained nature of the IIoT (and IoT) devices. Once a DDoS attack affects any device in the IIoT production line, it may cause the communication failure of the device and fail to connect to other devices in the IIoT system. The setting of interconnected IIoT leads to the immobility or loss of connection of the automated production line and leads to a system failure at scale. 


It is noted that DDoS attacks against the IIoT can cause unimaginable consequences. The following two cases illustrate the impact of DDoS attacks in large-scale IIoT systems~\cite{patil2021distributed}:

\begin{itemize}
    \item In February 2018, the target of a DDoS attack was GitHub, a popular online code managing service used by millions of developers. At this peak, the attack transmitted traffic at a rate of 1.3 Terabytes per second (TBps) and sent data packets at a rate of 126.9 million per second. The attacker benefits from the amplification consequence of a widespread database storing system called Memcached. By overflowing the Memcached server with misleading requests, the attacker was able to scale his attack by about 50,000 times.
    
    \item Another noted cyber-attacks on manufacturing systems was the December 2015 attack on the Ukrainian power grid. The attack combined multiple hacking techniques, including DDoS attacks, malicious emails, internet virus worms, and malware. The attackers managed to damage the whole power distribution company and caused more than 200,000 people in Ukraine to lose power for a certain period.
    
\end{itemize}




\section{A Reference Security Framework}
In this section, we discuss an overview of the proposed security framework against DDoS attacks for IIoT systems. In contrast to the IIoT framework mentioned in Section~\ref{fog-iiot-concept}, the concept of the `regional aggregator' has been added to the framework. A regional aggregator can seamlessly execute distributed services on a single device. The structure of the proposed IIoT framework is composed of four layers, as the architecture discussed in Section~\ref{fog-iiot-concept}. Recall, in this article, we try to take advantage of the underlying fog layers to mitigate DDoS attacks for the IIoT systems~\cite{kumar2021design}. In Fig.~\ref{fig-3-view} we illustrate a layered framework framework, which is composed of the following layers:

\begin{figure*}[t]
 \centering
    \includegraphics[scale=1.75]{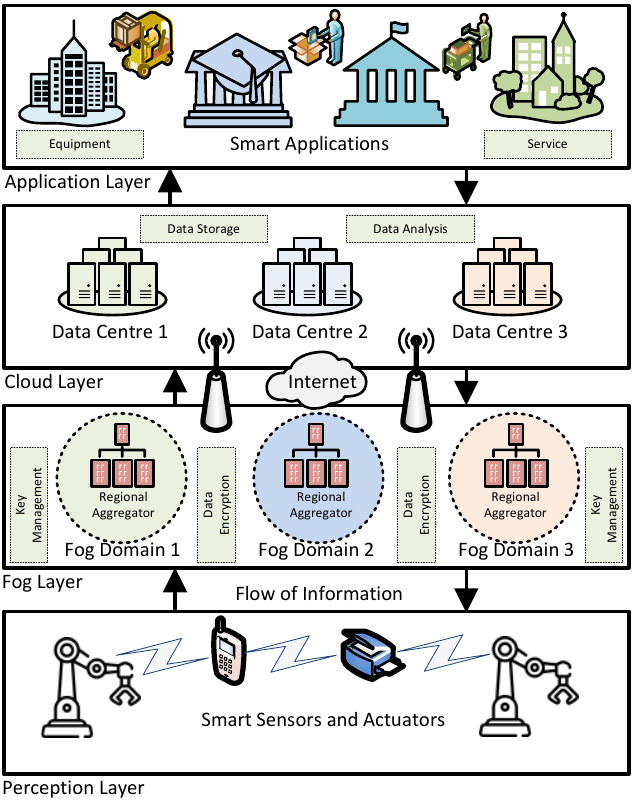}
    \caption{The proposed reference fog-based IIoT security framework.}
    \label{fig-3-view}
    \small
\end{figure*}

\begin{itemize}
    \item \textit{\textbf{Perception Layer:}} The perception layer is the bottom layer of the framework. It contains smart sensor devices, which perform data transmission and security authentication to the corresponding data collection.
    
    \item \textit{\textbf{Fog Layer:}} The second layer is the fog layer. In contrast to other IIoT fog layers, our framework introduces the concept of adding regional aggregate processors. The framework assumes a structure where each firewall is responsible for managing only a fixed number of devices in a fixed area. Each firewall uses RSA encryption for authentication between the firewalls. It ensures the accuracy of data transmission and allows for quick identification of the area under attack in the event of a cyber-attack, thus narrowing the scope and improving remediation efficiency.
    
    \item \textit{\textbf{Cloud Layer:}} Cloud layer is the third layer in the IIoT system. This layer is responsible for processing information and data storage in the cloud server and then distributing the data to the related devices and applications. The cloud layer is generally the same as secure storage, which could set the link between the application and fog layers.
    
    \item \textit{\textbf{Application Layer:}} It is the top layer of the architecture. This layer contains users and smart applications to access the desired services. The functionality of this layer is to deliver applications, e.g., smart cities, smart transportation, and smart agriculture, just a name a few applications. In addition, this layer links the smart devices or smart user's surface that offer smart services to the IIoT system's users. 
\end{itemize}

In the data-driven manufacturing environment, e.g., the IIoT, how to design a secure and distributed automation architecture has become a challenging task for technicians and designers. We argue that the balance control between fog computing and cloud computing will enable the IIoT to give more flexible and granular modularity to its applications and services at their highest potential. With this, many manufacturing organizations will benefit from automated systems that deploy sensors to measure, monitor, and analyze data through the IIoT to improve efficiency and increase revenue from operations. The volumes of data (e.g., big data) generated by these newly connected factories can be measured in petabytes - millions of streams of data from sensors connected to ICS (Industrial Control Systems) for both Operational Technology (OT) and Information Technology (IT) sides, SCADA (Supervisory Control and Data Acquisition) systems, and autonomous unattended machines, industrial robots effectively~\cite{falco2018iiot}.

Larger autonomous systems are now more likely to adopt new architectural approaches and solutions that enable IIoT to realize its extreme potential through the fog layer (sometimes referred to as edge computing). Fog computing is designed for data-intensive computational tasks that can be performed inside the local devices without the need for requiring high-performance capability. Fog can be seen as a developing distributed architecture that connects the cloud to edge devices in the production line, eliminating the need for a fixed Internet connection in the field. By selectively shifting computation, storage, communication and control, fog computing allows decisions to be made close to the edge sensors and actuators where the data is generated and used locally. That is close to the edge nodes where the actual data resides. The involvement of fog is a valuable complement to cloud computing, rather than a complete replacement so that IIoT can be used its advantages more efficiently, economically, safely and constructively in large-scale manufacturing environments.

As stated above, the fog and edge layers can be seen interchangeably, but there are key differences between them. The fog layer can be seen as a superset of the edge functions of the cloud layer. The architecture of the fog layer is designed primarily to combine resources and data sources with layers that reside at the cloud-to-edge, function-to-function or point-to-point levels to gain maximum efficiency at the edge level. In other words, edge computing is often limited to a small number of cloud-to-edge layers, often associated with simple protocol gateway functions. Concerning the IIoT systems, fog can provide better utilization of resources that can be processed locally and easily mitigate DDoS attacks at the edge level. This will, in turn, enable the IIoT system to detect and take appropriate measures to prevent such an attack at the edge level~\cite{kumar2021tp2sf}. 

The gateway nodes (e.g., a physical device or virtual platform) of the fog layer are the basic elements of the fog architecture. A fog node can be any device that provides the compute, network, storage, and acceleration elements of a fog architecture. Examples include industrial controllers, switches, routers, embedded servers, complex gateways, programmable logic controllers (PLCs), and smart IoT nodes (e.g., video surveillance cameras) used in IIoT automated production lines. The fog nodes result in the more reliable and faster data processing. Factories that can make use of the data flow from these fog node layers can not only make the devices more efficient but also interconnect data between factories. Fog nodes at lower levels of the overall IIoT architecture, e.g., individual computers, can be connected directly to local sensors and actuators to enable real-time data analysis and interpretation of anomalies. It can also respond and compensate for problems or solve them autonomously if authorised. In addition, a fog node can send appropriate service requests to providers with better technical resources, machine learning capabilities or maintenance services to obtain higher fog tiers.

When the environment in which they are located requires real-time decisions about data processing, e.g., raising an alert in case of equipment failure or adjusting parameter values at critical points in industrial production, fog nodes can provide millisecond delays for the process of data analysis and manipulation. This feature makes it unnecessary for plants to route such real-time decisions through cloud computing data centres. Helping to avoid potentially high latency events, queue delays, or network/server downtime could lead to engineering incidents, reduced productivity, or lower product quality. In addition, higher-level fog nodes in the factory allow for a broader view of industrial processes. Therefore, they can add extra functions, e.g., visualization of production line operations, monitoring the status of faulty machines, adjusting production strictures, modifying production schedules, ordering supplies, and sending alerts to the right people~\cite{abdali2021fog}.
Adding a fog layer and placing local fog nodes in the hierarchy near the IIoT pipeline, allowing them to be connected to sensors and actuators with cheap and fast local network facilities. Fog nodes improve safety and reduce the chance of leaks. If authorized, they can respond to abnormal conditions in milliseconds and quickly close the valve, significantly reducing the severity of spills. Fog nodes can operate between wired, fibre optic, and wireless networks and within those networks, making them best suited for connecting to industrial components based on the SCADA systems. Fog node analysis on local sites reduces the need for cloud bandwidth and overall costs. A balance of control between cloud and fog computing yields better results across the entire business process (cost, control, and security). Moving most decision-making functions to fog nodes and occasionally using the cloud to report status or receive commands or updates can help organizations create a superior control system.

\section{Discussion}
\label{discussion}
The significant finding of this article is how the overall security of the IIoT environment can be effectively enhanced by building a fog layer framework and strengthening the fog layer functionality to the core mitigation of DDoS attacks in an IIoT system~\cite{kumar2021toward}~\cite{alzoubi2021fog}. 


\subsection{Mitigation Considerations}

\begin{itemize}
    \item \textbf{\textit{Latency:}} Due to the dynamic nature of IIoT systems, more and more ICS require end-to-end data transfer delays in production lines to be kept to sub-milliseconds. However, this requirement is hard to be met by the leading providers of cloud services at present. In some cases, factories, the latency of up to 10 micro-seconds may be required to prevent production line shutdowns, avoid accidents, restore electrical service, or correct manufacturing errors. The creation of a fog node reduces latency because it eliminates the back-and-forth time delay from the sensor on the production line to the cloud and from the cloud back to the sensor. This latency includes transmission delays to the cloud (wireless or fibre optic devices), delays in waiting for data queues and delays in processing data by cloud servers, which can exceed 100 milli-seconds round trip even in a well-designed cloud network. We argue that using our framework, the local fog nodes can be able to react to conditions and make decisions within milli-seconds.
   
    \item \textbf{\textit{Security:}} Traditional industrial security measures focus on providing peripheral-based threat detection, and defences are not enough for IIoT systems. With fog computing, local security functionalities can be used in a more controlled and processed way for an organization. In general, fog nodes include a trusted hardware root, the foundation of a chain of trust that runs from the lowest sensors and actuators up the fog hierarchy to the cloud. Communications from the Internet to the distributed fog networks are monitored, and machine learning can be used to detect irregular activity in the local environment, allowing potential attacks to be detected and controlled in a timely manner. In future, blockchain technology can be an alternative to provide more secure storage that can easily be integrated with the proposed architecture.
    
    \item \textbf{\textit{Cognition:}} The fog layer structure can control the optimal location of the cloud to the edge computing, storage, and control functions on individual edge devices. The fog layer can also be employed to make data processing decisions via nearby fog nodes on many edge devices to avoid excessive data consumption and reduce transmission processing efficiency by transferring data to the cloud locally. The use of a fog layer allows data to be processed and analyzed easily and efficiently, autonomously in the local fog layer. The processed information is then sent through the fog layer to a data centre in the cloud for analysis. The addition of the fog layer enables the data processing performance of the IIoT system to support long-term planning and continuous system improvement in the future by monitoring local computation and processing at the edge devices.
    
    \item \textbf{\textit{Flexibility:}} A common industrial production environment may result in excessive fluctuations in data output. By sequencing all fog nodes, the fog hierarchy can be empowered to manage some of the unpredictable requirements, e.g., zero-day attacks of the system and allocate bearers beyond compliance to other under-utilized machines providing flexibility to the attack mitigation. The hierarchy of fog nodes can form a dynamic group to exchange information and enable an efficient federation of the learning environment. For example, suppose an industrial plant discovers in advance that it is running short of capacity for its tasks. It can then transfer part of its production tasks to another organization's assets and equipment for operation. That said, it can be easily performed with more flexibility than traditional approaches. The ability to change tasks dynamically also helps to coordinate and control information across the devices. 
    
    \item \textbf{\textit{Real-World Implementation:}} It is a key factor in the operational efficiency in real-world of the entire IIoT system. With the advent of Industry 4.0 and the evolution of Information and Communications Technology (ICT), from purpose-built and stand-alone systems to software-defined and modular interconnected operations, technicians are becoming aware that the cloud layer is gradually less productive to meet the need for efficiency. It is due to the complexity of connected edge systems and sensors and the integration of the various communication protocols and communication technologies. Fog nodes can, therefore, be added as an extension to traditional solutions or turned into an efficient protocol gateway by leveraging the computational efficiency of the edge devices in real-time. Recall that fog nodes collect and process data in different forms and use various protocols through the interconnection between the devices and the entire production pipeline so that devices and systems can be efficiently and uniformly connected without the need for different access methods for each component.
\end{itemize}

\subsection{Security Analysis and Lessons Learned}
The ability to analyze and process data at scale in real-time is considered to be one of the benefits of IIoT systems. In practice, the area of research on how to unify the data collected from various sources has been a potential challenge for individual developers, organizations, and users. This is because the devices may use different operating systems and complex communication protocols when sharing data. This is even more significant when considering the cross-domain data sharing between the nodes spread access to multiple jurisdictions. In addition, end-users are concerned about data security, and there are potential risks that industrial plants about sharing the sensitive information collected by the equipment with third parties outside the plant. We argue that fog architecture is beneficial where the locally computational performance can lead towards a direction where the various organizations can process such risks and their potential mitigation locally, particularly in avoiding DDoS attacks. 

It is noted that the advent of fog architecture provides an interconnected layer to the overall IIoT systems. From the security aspect, it is to ensure the IIoT systems can communicate and operate safely and efficiently between the equipment and the cloud layer, as well as data processing and usage. Further, a fog architecture can help identify component failures before data aggregation, ensuring the data analysis security, reliability, confidentiality, integrity, and availability. Since the fog layer is closer to the perception layer, which involves a large scale of smart sensors, fog nodes can identify and process data faster. This helps increase the efficiency of device authorization and data responsibility speed and problem-solving speed with high efficiency.

In addition, in the fog layer, when analyzing data received, the fog nodes could detect whether the system should enter the maintenance period or not. The fog layer could generate signs that a fault has occurred in the system, immediately alert the operator, carry out equipment inspections, and adjust the production line schedule, thus minimizing the impact of the hazard on the manufacturing chain. Moreover, there may also be unconventional situations where uncertainty may cause due to the errors in sensor readings signal an imminent failure, the probability of which might be minimal and allows the system to fix it before the dreaded failure occurs.

To maintain security, privacy, and trust in the IIoT system, integrity and availability of data usage are essential. Therefore, the fog layer needs to determine the content of the data and the importance of encryption in a hierarchical manner, sending only the appropriate data to the required processor terminals with the appropriate applicability. The current mainstream approach uses data classification, encryption, and virtual private networks (VPNs) to provide more secure communications. Those methods significantly reduce the risk of inadvertent cross-leakage of exclusive information in some directions.

\section{Conclusion and Future Work}

Compared to traditional IoT, IIoT involves a more extensive system of connected devices and nodes, resulting in more data processing tasks, ensuring that this data can be processed and transmitted safely and efficiently. Since the IIoT system involves large-scale devices and data process demands, the cloud computing system currently used in the IoT can no longer meet the data processing needs of large IIoT. Fog layer computing is proposed to extend cloud computing to edge IIoT devices. This article presents a conceptual framework integrating the fog layer to mitigate DDoS attacks in IIoT systems. However, there are still several open questions and challenges related to authentication and trust issues, Energy consumption, location awareness, etc., which need to be addressed for a more secure and safe fog-based IIoT architecture.

\ifCLASSOPTIONcaptionsoff
  \newpage
\fi

\bibliographystyle{IEEEtran}
\bibliography{bare_jrnl}

\end{document}